\documentstyle{article}\begin{document}
\title{Quantum scalar field propagator in a stochastic gravitational plane wave}

\author{Z. Haba \\
Institute of Theoretical Physics, University of Wroclaw,\\ 50-204
Wroclaw, Plac Maxa Borna 9,
Poland\\email:zbigniew.haba@uwr.edu.pl}
\date{\today}
\maketitle
\begin{abstract} A stochastic metric can appear in classical as well as in quantum gravity.
We show that if the linearized stochastic Gaussian gravitational
plane wave has the frequency spectrum $\omega^{4\gamma-1}$ (
$0\leq\gamma<1$ ) then the equal-time propagator of the scalar
field behaves as $p^{-\frac{1}{1-\gamma}}$ for large momenta. We
discuss models of quantum field theory where such anomalous
behavior can appear.
\end{abstract}
\section{Introduction}The Standard Model on a stochastic background of gravitational waves
or the one supplemented with quantum gravity  can be viewed as a
theory of matter fields in a random geometry which can be singular
at short  distances.There are some results
\cite{deser}\cite{habaplb}\cite{jurk}\cite{horava}\cite{reuter}\cite{carlip}\cite{carlip2}
indicating  that the gravity can modify the short distance
(equivalently large momentum) behavior of matter fields. There are
some simplified Lorentz non-invariant models where such a modified
short distance behavior in various directions in space-time is
realized \cite{horava2}\cite{verlinde}\cite{kabat}. The question
appears whether the large momentum behavior of quantum fields in
the background of quantized gravity or in the stochastic
background could be experimentally verified, e.g., in high energy
scattering. The detection of gravitational waves \cite{detection}
raises the questions whether these waves can be considered as a
stream of gravitons and whether their interaction with particles
can be treated as a particle-graviton scattering. There is a
suggestion \cite{wilczek} that the effect of gravitons can be
observed as a noise in the interferometers applied for
gravitational waves detection. In such a model the quantum
equation for geodesic deviation is studied. It is shown that
gravitons transform this equation into a stochastic differential
equation changing the particle evolution.

     In this paper we are interested in the problem how
     the dynamics of the quantum scalar field can be changed  in a linearly polarized
stochastic gravitational plane wave moving in the $x$-direction.
Such a stochastic metric can arise from stochastic sources of
gravitational waves or from particular states of quantum gravity.
The stochastic metric corresponding to the plane wave has the
spectral function $\rho({\bf k})$ restricted to $k_{x}\geq 0$ with
$k_{y}=k_{z}=0$.We show that if the graviton correlation function
is singular at short distances then the scalar field propagator
decays faster in the momentum space (in comparison to the free
field propagator)  in the direction orthogonal to the wave
propagation. Such an improved behavior of propagators of matter
fields in the background of quantum gravitational field is
important for renormalization and ultraviolet stability of these
theories. The singular correlation functions of the metric (large
contribution of high frequencies) come out as a consequence of the
perturbative quantization of gravity. The high frequency
dependence of the gravitational radiation can appear as a result
of an inflationary enhancement of the gravitational vacuum
contribution \cite{kuchar}\cite{hartle} according to
ref.\cite{japan}. The plan of the paper is the following. In sec.2
we discuss the Gaussian stochastic plane waves and their
interaction with a quantum scalar field. In sec.3 we consider
quantum states of the linearized gravitational field which can
describe a stochastic metric. In sec.4 the Dyson expansion for the
propagator in a simplified model of a random metric is
investigated. In sec.5 we consider the Feynman path integral
representing the resummation of the Dyson series. In sec.6 we
estimate the behavior of the propagator at large momenta. In sec.7
we summarize our results.

\section{Quantum scalar field in a  stochastic plane wave}
We consider a perurbation $h_{\mu\nu}(\xi)$ ( $\xi=(t,{\bf x})$)
of the Minkowski metric
 in the traceless transverse (TT) gauge requiring in
addition that the metric tensor is diagonal and assuming that
$h_{\mu\nu}$ is propagating only along the $x$ axis. Then, the
metric is
\begin{equation}\begin{array}{l}
ds^{2}=g_{\mu\nu}d\xi^{\mu}d\xi^{\nu}\equiv dt^{2}-(
\delta_{jk}+h_{jk})dx^{j}dx^{k}\cr\equiv dt^{2}- dx^{2}-(1-
h(t,x))dy^{2}-(1+h(t,x))dz^{2}.\end{array}
\end{equation}
Such a metric can describe a gravitational wave moving along the
$x$-axis. It can be considered as a solution of the
$TT$-conditions $\partial_{j}h_{jk}=0$ and $h^{j}_{j}=0$. For the
plane wave  moving in the positive direction
of the $x$-axis \cite{inv} $h(t,x)=h(u)$ where $u=t-x$ (the velocity of
light $c=1$) .

 The
propagator of the scalar field (with mass $m$) in a stochastic
metric is defined by a functional integral
\begin{equation}\begin{array}{l}
G(\xi,\xi^{\prime})=Z^{-1}\Big<\int {\cal
D}\phi\exp\Big(\frac{i}{2\hbar}\int d^{4}x\sqrt{\vert
g\vert}(g^{\mu\nu}\partial_{\mu}\phi\partial_{\nu}\phi -
m^{2}\phi^{2}\Big)\phi(x)\phi(y)\Big>\cr
=\frac{1}{2}i\hbar\Big<{\cal A}^{-1}(\xi;\xi^{\prime})\Big>,
\end{array}\end{equation} where $Z$ is a
normalization factor, $g=\det(g_{\mu\nu}) $ is the determinant of
the metric $g_{\mu\nu}$ and ${\cal A}^{-1}(x,y)$ denotes the
kernel (Green function) of the inverse of the operator defined by
the bilinear form in the Lagrangian of the scalar field in eq.(2)
\begin{equation}
2{\cal A}=\vert g\vert^{-\frac{1}{2}}\partial_{\mu}\vert
g\vert^{\frac{1}{2}}g^{\mu\nu}\partial_{\nu}+m^{2}
\end{equation}
In eq.(2) the metric is understood  as an average over a
stochastic background of gravitational waves \cite{gw} which may
have a primordial origin (quantum states of the gravitational
field, sec.3) or may come from countless sources emitting
gravitational waves in the universe (e.g., from merging of
primordial black holes). The average (2) could also be performed
in the Feynman path integral expressing the quantum fluctuations
of the metric as discussed in
\cite{ford1}\cite{ford2}\cite{ford3}. We assume that $h$ can be
approximated by a Gaussian variable. Such an assumption, in the
case of a stochastic background, is justified by the central limit
theorem of probability theory. If we have small stochastic sources
of gravitational radiation then the probability distribution of an
infinite sum of stochastic variables (no matter what are their
individual probability distributions) can be approximated by  the
Gaussian distribution. In quantum theory of the next section we
show that in the limit of linearized semiclassical gravity we can
use the stochastic Gaussian approximation for quantum plane waves.
\section{Linearized quantum gravity} We consider a model of the
quantum scalar field interacting with linearized quantum gravity
described by a small perturbation $h_{\mu\nu}$ of the Minkowski
metric. In the Hamiltonian framework of this theory we have the
Schr\"odinger equation \cite{kuchar}
\begin{equation}
i\hbar\partial_{t}\psi=(H_{gr}(h)+H_{sc}(h,\phi))\psi,
\end{equation}
where $H_{gr}$ is the Hamiltonian for the linearized gravity (the
gravitons) and $H_{sc}$ is the Hamiltonian of the quantum scalar
field in an external metric $h$. We discuss  the correlation
functions \begin{equation} (U_{t}\psi_{0},\phi({\bf
x})U_{t}\phi({\bf x}^{\prime})\psi_{0}) =(\psi_{t},\phi({\bf
x})\phi(t,{\bf x}^{\prime})\psi_{t}),
\end{equation}where $U_{t}$ is the unitary evolution generated by the
Hamiltonian (4) and $\phi(t,{\bf x})=U_{t}\phi({\bf x})U_{t}^{+}$.
We can write
\begin{equation}
U_{t}=\exp(-\frac{i}{\hbar}H_{gr}t)T\Big(\exp(-\frac{i}{\hbar}\int_{0}^{t}H^{I}_{sc}(s)ds)\Big)\equiv
U^{gr}(t)U_{sc}^{I}(t),
\end{equation}
where $T(...)$ denotes the time-ordered exponential,
$U^{gr}(t)=\exp(-\frac{i}{\hbar}H^{gr}t)$ and
\begin{displaymath}
H^{I}_{sc}(s)=U^{gr}(s)^{+}H_{sc}(h,\phi)U^{gr}(s)=H_{sc}(h(s),\phi)
\end{displaymath}
with\begin{equation} h(s,{\bf
x})=\exp(\frac{i}{\hbar}H_{gr}s)h({\bf
x})\exp(-\frac{i}{\hbar}H_{gr}s).
\end{equation}
We assume that the initial state is of the product form
$\psi_{0}(h,\phi)=\psi_{0}^{gr}(h)\psi^{sc}_{0}(\phi)$. Then
 \begin{equation}\begin{array}{l}
U_{t}\psi_{0}^{gr}(h)\psi^{sc}_{0}(\phi)\cr=
\Big(\exp(-\frac{i}{\hbar}H_{gr}t)\psi_{0}^{gr}(h)\Big)
T\Big(\exp(-\frac{i}{\hbar}\int_{0}^{t}H^{I}_{sc}(s)ds)\Big)\psi_{0}^{sc}(\phi)
\cr\equiv \psi_{t}^{gr}(h)\psi_{t}^{sc}(h(.),\phi),
\end{array}\end{equation}where $\psi_{t}^{sc}(h(.),\phi)$ becomes a functional of the quantum
field $ h(s,{\bf x})$. So far the formulas are exact. $h(.)$ on
the rhs of eq.(8) means that $H_{sc}(h,\phi)\rightarrow
H_{sc}(h(s),\phi) $. We are using in $H_{sc}(h(s),\phi)$ the
scalar field Hamiltonian with an operator metric $h(s,{\bf x})$
(7). We make the approximation neglecting  the non-commutativity
of $h(s, {\bf x})$ at different times.  $h(s,{\bf x})$ becomes  a
classical stochastic field. The commutator $[h(s, {\bf x}),
h(s^{\prime}, {\bf x}^{\prime})]\simeq O(\hbar)$. This means that
neglecting non-commutativity we are applying a semiclassical
approximation to the coupling of the metric to the quantum scalar
field. We choose as $\psi_{0}^{sc}(\phi) $ the ground state of the
massive free scalar field. We apply the functional representation
of states in quantum field theory. In this representation the
propagator in a stochastic gravitational field of the quantum
scalar field in its ground state is

\begin{equation}\begin{array}{l}
(U_{t}\psi_{0},\phi({\bf x})\phi(t,{\bf
x}^{\prime})U_{t}\psi_{0})\cr= \int {\cal D}h_{rl}\vert
\psi^{gr}_{t}\vert^{2}{\cal D}\phi \exp\Big(\frac{i}{2\hbar}\int
d^{4}x\sqrt{\vert
g\vert}(g^{\mu\nu}\partial_{\mu}\phi\partial_{\nu}\phi -
m^{2}\phi^{2}\Big)\phi({\bf x})\phi(t,{\bf x^{\prime}})\cr
=\frac{1}{2}i\hbar\int {\cal D}h_{rl}\vert
\psi^{gr}_{t}(h)\vert^{2}{\cal A}^{-1}(0,{\bf x};t,{\bf
x}^{\prime}).
\end{array}\end{equation}
The Hamiltonian for the linearized (free) gravitational field is
\cite{kuchar}
\begin{equation}
H_{gr}=\frac{1}{2}\int d{\bf k}\Big(-\Lambda\frac{\delta}{\delta
h_{rl}({\bf k})}\Lambda\frac{\delta}{\delta h_{rl}(-{\bf
k})}+k^{2}(\Lambda h)_{rl}({\bf k})(\Lambda h)_{rl}(-{\bf
k})\Big),\end{equation} where $r,l=1,2,3$ and $(\Lambda T)_{ij}$
is  the projection of a tensor $T_{mn}$ onto the one in the TT
gauge defined at the beginning of sec.2. The matrix  $\Lambda$ in
the momentum representation has the form

\begin{displaymath}\begin{array}{l}
2\Lambda_{ij;mn}(\frac{{\bf k}}{k})=(\delta_{im}-k^{-2}k_{i}k_{m})
(\delta_{jn}-k^{-2}k_{j}k_{n})\cr+(\delta_{in}-k^{-2}k_{i}k_{n})
(\delta_{jm}-k^{-2}k_{j}k_{m})-\frac{2}{3}(\delta_{ij}-k^{-2}k_{i}k_{j})
(\delta_{nm}-k^{-2}k_{n}k_{m}).\end{array}
\end{displaymath}
 It follows from eqs.(7) and (10) that
\begin{equation}
(\partial_{t}^{2}-\triangle)h_{rl}=0.
\end{equation}
Moreover, as shown in \cite{kuchar} the correlation functions in
the ground state of the Hamiltonian (10) (as calculated below in
eq.(17)) satisfy the Weinberg's requirements for massless tensor
fields \cite{weinberg}.

 We choose as an initial state the general Gaussian translation invariant
 wave function of tensorial fields in the TT gauge
\begin{equation}
\psi_{0}^{gr}=A_{0}\exp\Big(-\frac{1}{2\hbar}\int (\Lambda
h)_{rl}({\bf x})\Gamma_{0}({\bf x}-{\bf y})_{rl,ij}(\Lambda
h)_{ij}({\bf y})d{\bf x}d{\bf y}\Big),
\end{equation}where  $A_{0}$ is a normalization constant.
The Gaussian states have a positively definite Wigner function.
For this reason they give a proper semiclassical approximation for
quantum states. For a free Hamiltonian (10) the time evolution of
a Gaussian state is again a Gaussian state. The Schr\"odinger
equation $i\hbar\partial_{t}\psi=H_{gr}\psi$ with the initial
condition (12) has the solution which is again a Gaussian
translation invariant wave function
\begin{equation}
\psi_{t}^{gr}=A_{t}\exp\Big(-\frac{1}{2\hbar}\int(\Lambda
h)_{rl}({\bf x})\Gamma_{t}({\bf x}-{\bf y})_{rl,ij}(\Lambda
h)_{ij}({\bf y})d{\bf x}d{\bf y}\Big),
\end{equation}
if $\Gamma$ satisfies the equation
\begin{equation}i\partial_{t}\Gamma_{t}({\bf k})_{rl,ij}-
\Gamma_{t}({\bf k})_{rl,mn}\Gamma_{t}(-{\bf
k})_{mn,ij}+k^{2}\delta_{ri}\delta_{lj}=0,
\end{equation}where $\Gamma({\bf k})$ is the Fourier transform of $\Gamma({\bf x})$ and $k=\vert{\bf
k}\vert$. The time-independent solution of eq.(14)
\begin{equation}
\Gamma_{t}({\bf k})_{rl,ij}=k\delta_{ri}\delta_{lj} \end{equation}
 corresponds to the ground state \cite{kuchar}\cite{hartle}

\begin{displaymath}
\psi_{0}^{gr}(h)=\exp\Big(-\frac{1}{2}\int d{\bf k}\vert {\bf
k}\vert (\Lambda h)_{jl}({\bf k})^{*}(\Lambda h)_{jl}({\bf
k})\Big)
\end{displaymath} leading to the Gaussian measure\begin{displaymath}
d\mu_{0}(h)={\cal D} h_{jk}\vert \psi_{0}^{gr}\vert^{2}
\end{displaymath}
describing in the TT gauge  the free tensorial massless  fields in
the ground state. Then, correlation functions in the ground state
are
\begin{equation}\begin{array}{l}
(\psi_{0}^{gr},h_{jl}({\bf x})h_{mn}(t,{\bf
x}^{\prime})\psi_{0}^{gr})=(2\pi)^{-3}\int d{\bf k}\vert {\bf
k}\vert^{-1}\exp(-ikt)\exp(i{\bf k}({\bf x}-{\bf x}^{\prime}))
\cr(\delta_{jm}({\bf k})\delta_{ln}({\bf k})+\delta_{jn}({\bf
k})\delta_{lm}({\bf k})-\delta_{jl}({\bf k})\delta_{mn}({\bf k})),
\end{array}\end{equation}where
\begin{displaymath}
\delta_{jl}({\bf k})=\delta_{jl}-k_{j}k_{l}\vert {\bf
k}\vert^{-2}.
\end{displaymath}
The correlation (16) behaves as
\begin{equation}
\Big((\xi-\xi^{\prime})^{2}\Big)^{-1}
=\Big((u-u^{\prime})(x-t-x^{\prime})-(y-y^{\prime})^{2}-(z-z^{\prime})^{2}\Big)^{-1}
\end{equation}
at short space-time distances (in the notation of sec.2).

The time-dependent solutions of eq.(14) can be related to
solutions of a linear equation. Let us define the matrix $v({\bf
k})$ as a time-ordered exponential of the matrix $\Gamma({\bf k})$
\begin{equation}
v=T\Big(\exp(i\int_{0}^{t}\Gamma_{s}ds)\Big).
\end{equation}
It can also be defined as the solution of the equation
$\partial_{t}v=i\Gamma v$. If $\Gamma$ satisfies eq.(14) then the
matrix $v$ satisfies a linear equation
\begin{equation} \frac{d^{2}v}{dt^{2}}+k^{2}v=0.
\end{equation}
We can recover $\Gamma$ from $v$ as
\begin{equation}i\Gamma_{t}=v^{-1}\frac{dv}{dt}.
\end{equation}
The general solution of eq.(19) is \begin{equation}
v=M\exp(ikt)+N\exp(-ikt),
\end{equation}
where the matrices $M$ and $N$ may depend on ${\bf k}$. We have
from eqs.(20)-(21) that \begin{equation}
\Gamma_{0}=k(M-N)(M+N)^{-1}
\end{equation}
and \begin{equation} \Gamma_{t}({\bf k})=k(\exp(ikt)M-N\exp(-ikt))
(\exp(ikt)M+N\exp(-ikt))^{-1}.
\end{equation}The two-point correlation function of  $h$
calculated as the covariance of the measure $d\mu={\cal D}h_{rl}
\vert \psi_{t}^{gr}\vert^{2}$ is the Fourier transform of
$\Gamma_{t}^{-1}({\bf k})$ . By a proper choice of the (matrix)
functions $M$ and $N$ we can achieve that the dependence of matrix
elements of $\Gamma^{-1}$ on the components of ${\bf k}$ can be
dominating or negligible. We can choose $\Gamma$ in eqs.(22)-(23)
in such a way that only the correlation functions
$<h_{33}h_{33}>=<h_{22}h_{22}>= -<h_{22}h_{33}> $ are not
negligible (we demand ($h_{33}=-h_{22}=h$ in the sense of
correlation functions). The remaining components of the metric in
the TT gauge $h_{11}$ and $h_{jk}$ for $j\neq k$ will have
vanishing correlation functions by a choice of $\Gamma$. We choose
$\Gamma_{rl,ij}$ such that $\Gamma_{rl,rl} \rightarrow \infty$ for
$rl=11,12,13,23$. In such a case the expectation values (and
correlation functions) with respect to the measure $d\mu={\cal
D}h_{rl} \vert \psi_{t}^{gr}\vert^{2}$ of the components of the
tensor  $h_{rl}$ which are absent in the metric (1)  tend to zero.
In this way the wave function $\psi_{t}^{gr}$ is imposing the TT
gauge in the sense that the correlation functions of $h$
asymptotically satisfy this gauge. The TT gauge (for diagonal $h$)
has as a consequence that $h$ depends only on $t$ and $x$. It
follows that the wave equation (11) takes the form
$(\partial_{t}^{2}-\partial_{x}^{2})h=0$ (in the sense of
expectation values) with the solution which is a sum of waves
depending either on $t-x$ or on $t+x$.
  We can
damp the $t+x$ component choosing $\Gamma_{22,22}({\bf
k})=\Gamma_{33,33}({\bf k})\rightarrow \infty $ when $k_{x}<0$.

Summarizing, we have shown in this section that  a gravitational
wave treated as a metric in a particular quantum state of a
linearized quantum gravity can be considered as a stochastic
Gaussian plane wave. We discuss a scalar quantum field in such a
stochastic wave in subsequent sections.

\section{Dyson expansion of the scalar field propagator}
We consider the metric (1).
 We assume that we have a small  perturbation of the Minkowski
metric, so that $h^{2}\simeq 0$. In such a case  the term
$\partial_{t}\ln \vert g\vert
\partial_{t}-\partial_{x}\ln \vert g\vert
\partial_{x}$
which is of the first order in derivatives for  the scalar wave
operator in eq.(3) is absent because $g=\det(g_{jk})=1-h^{2}\simeq
1 $. For the same reason we make the approximation $(1\pm
h)^{-1}\simeq 1\mp h$. If $h$ depends only on $u=t-x$ then we can
change the variable $t\rightarrow t-x$  in ${\cal A}$. Then, the
coefficients of the differential operator ${\cal A}$ in eq.(3)
depend on one variable $u$
\begin{equation}
{\cal A}=-\frac{1}{2}(\partial_{u}^{2}-\triangle +m^{2})
+\frac{1}{2}h(u)\partial_{y}^{2}-\frac{1}{2}h(u)\partial_{z}^{2}.
\end{equation}
We can  take the Fourier transform in spatial variables of the
operator (24) and its kernel
\begin{equation}
\tilde{\cal A}^{-1}(u,u^{\prime},{\bf p})=\int d{\bf x}\exp
(ip_{x}(x-x^{\prime})+ip_{y}(y-y^{\prime})+ip_{z}(z-z^{\prime})){\cal
A}^{-1}(\xi;\xi^{\prime})
\end{equation}
It follows that $ \tilde{\cal A}^{-1}(u,u^{\prime},{\bf p})$ is
the kernel of the operator
\begin{equation}
\tilde{{\cal A}}=-\frac{1}{2}(\partial_{u}^{2}+{\bf p}^{2} +m^{2})
+\frac{1}{2}h(u)p_{+}p_{-}\equiv\tilde{{\cal A}}_{0}+V
\end{equation}
acting in the space of functions $ \psi(u,{\bf p})$ (where the
action of the momentum ${\bf p}$ in $\tilde{{\cal A}}$ is just the
multiplication). Here, ${\bf p}^{2}=p_{x}^{2}+p_{y}^{2}+p_{z}^{2}$
, $p_{+}=p_{z}+p_{y}$,$p_{-}=p_{z}-p_{y}$,  \begin{equation}
\tilde{{\cal A}}_{0}=-\frac{1}{2}\partial_{u}^{2}-\frac{1}{2}{\bf
p}^{2}-\frac{1}{2}m^{2}
\end{equation}and
\begin{equation}V=\frac{1}{2}
h(u)p_{-}p_{+}.
\end{equation}
With minor changes in the discussion of the propagator (25) we can
consider the case of the metric $h$ depending only on  $x$ . This
changes $ \tilde{{\cal A}}_{0}$ in eq.(26) as
\begin{equation}
\tilde{{\cal
A}}_{0}=\frac{1}{2}\partial_{x}^{2}+\frac{1}{2}p_{0}^{2}-\frac{1}{2}{\bf
p}_{\perp}^{2}-\frac{1}{2}m^{2},
\end{equation}where \begin{equation}{\bf
p}_{\perp}^{2}=p_{y}^{2}+p_{z}^{2} .\end{equation} The form of the
operator $V$ (28) does not change except that $h$ depends only on
$x$.

We shall consider a random metric $h$. This randomness can come
either from classical sources as discussed in sec.2 or from
quantum fluctuations of sec.3. For explicit calculations we need a
dependence of $h$ solely on one coordinate. There can be a
physical reason for such an approximation. In the case of the
classical waves this requirement relies on the assumption that we
deal with plane waves. In the quantum realm such a metric
depending  solely on $t-x$ or on $x$ can arise from a particular
initial state as discussed in sec.3.

 We perform the calculations for the operator $\tilde{{\cal
A}}$ (26) with $\tilde{{\cal A}}_{0}$ (27) corresponding to the
plane wave. The calculations with $\tilde{{\cal A}}_{0}$ (29) are
similar (just replace $\tilde{{\cal A}}$ by $-\tilde{{\cal A}} $
and $u $ by $x$). We represent the propagator $\tilde{{\cal
A}}^{-1}=i\int_{0}^{\infty}d\tau \exp(-i\tau \tilde{{\cal A}})$ in
terms of the proper time Hamiltonian evolution. The operator
$\tilde{{\cal A}}$ has the form of a Hamiltonian in quantum
mechanics. We assume that it has a self-adjoint extension. The
unitary evolution $\exp(-i\tilde{{\cal A}}\tau)$ can be expressed
in the interaction picture.  We write
\begin{equation} U_{\tau}=\exp(-i\tilde{{\cal
A}}\tau)=\exp(-i\tilde{{\cal A}}_{0}\tau)U^{I}_{\tau}.
\end{equation}
Then,
\begin{equation}
\partial_{\tau}U_{\tau}^{I}=-iV_{\tau}U_{\tau}^{I},
\end{equation}
where
\begin{equation}
V_{\tau}=\exp(i\tilde{{\cal A}}_{0}\tau)V\exp(-i\tilde{{\cal
A}}_{0}\tau).
\end{equation}We are interested in an average over the random field $h$.
We can calculate such averages expanding the solution of eq.(32)
in the Dyson series
\begin{equation}
U_{\tau}=\exp(-i\tilde{{\cal
A}}_{0}\tau)U_{\tau}^{I}=\exp(-i\tilde{{\cal
A}}_{0}\tau)\Big(1-i\int_{0}^{\tau}dsV_{s}-\int_{0}^{\tau}ds_{2}\int_{0}^{s_{2}}ds_{1}V_{s_{2}}V_{s_{1}}+...\Big).
\end{equation}
The expectation value over the metric  in the lowest order is
\begin{equation}\begin{array}{l}
<(U_{\tau}\psi)(u)>=(\exp(-i\tau
H_{0})\psi)(u)\cr-\int_{0}^{\tau}ds_{2} \int_{-\infty}^{\infty}
du_{1}\int_{-\infty}^{\infty}du_{2}(\exp(-i(\tau-s_{2})H_{0}))(u,u_{1})
<V(u_{2})V(u_{1})>\cr\times
(\exp(-i(s_{2}-s_{1})H_{0}))(u_{1},u_{2})\int_{0}^{s_{2}}ds_{1}
(\exp(-is_{1}H_{0}))(u_{2},u_{3})\psi(u_{3})du_{3} ,
\end{array}\end{equation} where
\begin{equation}\begin{array}{l}
(\exp(-is\tilde{{\cal A}}_{0}))(u,u^{\prime})=(2i\pi
s)^{-\frac{1}{2}} \exp\Big( \frac{i}{2s}\vert
u-u^{\prime}\vert^{2}\Big) \exp(i\frac{s}{2}(m^{2}+{\bf
p}^{2}))\cr\equiv \exp(i\frac{s}{2}(m^{2}+{\bf
p}^{2}))p(s;u-u^{\prime}).
\end{array}\end{equation}
Going from eq.(34) to eq.(35) we have changed the proper time
integration variables $s_{1}\rightarrow \tau -s_{2}$ and $
s_{2}\rightarrow \tau-s_{1}$.
\section{Feynman path integral representation}
For a  non-perturbative averaging over $h$  it is useful to
express the evolution $U_{\tau}$ by means of the Feynman path
integral
\begin{equation}\begin{array}{l}
(U_{\tau}\psi)(u)=\exp(i\tau\frac{1}{2}({\bf
p}^{2}+m^{2}))\int_{q(0)=u} {\cal D}
q(.)\cr\times\exp\Big(\frac{i}{2}\int_{0}^{\tau}ds(\frac{dq}{ds})^{2}\Big)\exp(-i\int_{0}^{\tau}dsV(
q(s)))\psi( q(\tau)).
\end{array}\end{equation}
The mean value  of the evolution operator in the Gaussian  random
metric can be evaluated as
\begin{equation}\begin{array}{l}
<(U_{\tau}\psi)(u)>=\exp(i\tau\frac{1}{2}({\bf
p}^{2}+m^{2}))\int_{q(0)=u} {\cal
D}q(.)\exp\Big(\frac{i}{2}\int_{0}^{\tau}(\frac{dq}{ds})^{2}\Big)
\cr\exp(-\frac{1}{2}\int_{0}^{\tau}ds\int_{0}^{\tau}ds^{\prime}<V(q(s))V(q(s^{\prime}))>)\psi(q(\tau)).
\end{array}\end{equation}The Feynman path integral (37) can be considered as a resummation of the Dyson series (34).
 The lowest order term in the expansion of the
Feynman formula for $U_{\tau}\psi$ in eq.(37) is
\begin{equation}\begin{array}{l}
<(U_{\tau}\psi)(u)>=\exp(i\tau\frac{1}{2}({\bf
p}^{2}+m^{2}))\int_{q(0)=u} {\cal
D}q(.)\exp\Big(\frac{i}{2}\int_{0}^{\tau}(\frac{dq}{ds})^{2}\Big)\cr
\times
\Big(1-\int_{0}^{\tau}ds_{2}\int_{0}^{s_{2}}ds_{1}<V_{s_{2}}V_{s_{1}}>+...\Big)\psi(q(\tau)).
\end{array}\end{equation}
It agrees with the Dyson expansion (35). The equality of the
expansion in $V$ of the Feynman integral (38)-(39) and the Dyson
expansion (34)-(35) can be shown using the expression for the
calculation of the Feynman integral of "cylinder functions"
\cite{ginibre}\cite{simon}. If $0\leq s_{1}\leq s_{2}\leq.....\leq
s_{n-1}\leq \tau$ then  the Feynman integral of functions of paths
starting from $q(0)=u$ and depending on a finite set of points
(cylinder functions) $(q(s_{1}),q(s_{2}),....,q(s_{n-1}),q(\tau))$
is
\begin{equation}\begin{array}{l}
\int_{q(0)=u} {\cal
D}q(.)\exp\Big(\frac{i}{2}\int_{0}^{\tau}(\frac{dq}{ds})^{2}\Big)
F(q(s_{1}),q(s_{2}),....,q(s_{n-1}),q(\tau))\cr =\int
p(s_{1},u_{1}-u)p(s_{2}-s_{1};u_{2}-u_{1})....p(\tau-s_{n-1};u_{n}-u_{n-1})
\cr\times
F(u_{1},...,u_{n})du_{1}...du_{n}\end{array}\end{equation} for any
function $F(u_{1},...,u_{n})$ of $n$-variables. Eqs.(38)-(39)
coincide with eq.(35) if in the $(n-1)$th order of Dyson
perturbation expansion we change the proper time integration
variables $s_{j}\rightarrow \tau-s_{n-j}$ (as we did in eq.(35)
for $n=3$).

 We may express the time evolution in terms of the evolution
 kernel $K_{\tau}$
\begin{equation}
(U_{\tau}\psi)(u)=\int du^{\prime}K_{\tau}(u,u^{\prime};{\bf
p})\psi(u^{\prime}).
\end{equation}
The two-point function for the scalar field is obtained as
 the resolvent
of $\tilde{{\cal A}}$
\begin{displaymath}
G(\tilde{{\cal A}})= i\int_{0}^{\infty}d\tau\exp(-i\tilde{{\cal
A}}\tau)=\tilde{{\cal A}}^{-1}.\end{displaymath} Its kernel is (in
time coordinates $u$ and in spatial momenta)

\begin{equation}\begin{array}{l}
G(\tilde{{\cal A}},u,u^{\prime};{\bf p})=
i\int_{0}^{\infty}d\tau\exp(-i\tilde{{\cal
A}}\tau)(u,u^{\prime})=i\int_{0}^{\infty}d\tau
K_{\tau}(u,u^{\prime};{\bf p}).\end{array}\end{equation} As
discussed in secs.2 and 3 we impose on the Gaussian measure
$d\mu(h)= {\cal D}h_{rl} \vert \psi_{t}^{gr}\vert^{2}$ the
conditions that in the linearized quantum gravity the measure is
strongly concentrated on the metric solving the wave equation and
satisfying the TT condition. In such a case (as discussed at the
end of sec.3) the metric depends only on $u=t-x$. The background
of plane gravitational waves discussed in sec.2 moving in a fixed
direction can also depend on one variable. In these idealized
cases we assume that $h(u)$ is a Gaussian random field defined by
the measure $d\mu(h)$ characterized by its generating functional
\begin{equation}
\int d\mu(h)\exp\Big(i\int du J(u)h(u)\Big)=\exp\Big(-\frac{1}{2}
\int dudu^{\prime}J(u)g(u-u^{\prime})J(u^{\prime})\Big),
\end{equation}
where $g(u-u^{\prime})$ is the correlation function of $h(u)$.
 We consider the covariance
\begin{equation}
g(u-u^{\prime})=<h(u)h(u^{\prime})>=\kappa^{2}\vert
u-u^{\prime}\vert^{-4\gamma}.
\end{equation}
We need $0<\gamma<\frac{1}{4}$ for a rigorous approach (the
restriction $\gamma<\frac{1}{4}$ and consequences of  crossing the
line $\gamma=\frac{1}{4}$ are considered at the end of sec.6). The
exact scale invariance of $g(u)$ is not necessary for our results.
It is sufficient to assume that $g(u)$ has the form (44) for $
u\rightarrow 0$. We make the assumption of an exact scale
invariance for simplicity of the argument. The absolute value
$\vert u-u^{\prime}\vert$  in eq.(44) is needed for a mathematical
definition of a stochastic variable as the bilinear form in the
exponential in eq.(44) is positively definite only with the
absolute value and $0<\gamma<\frac{1}{4}$. However, in the next
section only the scaling property of the covariance (44) is used
in a derivation of the result. Let us note that the free field
correlation function (17) behaves as $(u-u^{\prime})^{-1}$ (so
$\gamma=\frac{1}{4}$) if $y=y^{\prime}=z=z^{\prime}=0$. In the
free field correlation (17) there is no absolute value but instead
the $ i\epsilon $ prescription for an interpretation of the
integral over $u,u^{\prime}$ in the Feynman propagator. We set
$\gamma$ as a free parameter
 taking into account the suggestions
\cite{deser}\cite{habaplb}\cite{jurk}\cite{carlip}\cite{horava}
that quantum gravity at short distances can be more regular than
in the canonical field theory. For a stochastic background of
gravitational waves the scale invariant correlation functions (44)
have been derived in some models of gravitational radiation
\cite{gw1}\cite{gw2}.
 It is our aim in this paper to show
that singular gravity leads to a more regular scalar field
propagator.
  The
singularity of the correlations  at small time means a large
contribution of high-frequency waves to the spectrum  (as the
frequency spectrum of the covariance (44) is $\omega^{4\gamma-1}$)
. At present only low frequency gravitational waves have been
detected \cite{gw}. The large contribution of high frequency modes
to the wave spectrum is expected to have a quantum origin. These
quantum fluctuations could be enhanced by inflation and appear in
observable wave spectrum as discussed in \cite{japan}(a similar
mechanism is known to work for scalar fields leading to the
squeezed states \cite{sidorov}\cite{albrecht}).

\section{Estimates on the propagator}

We  have for a Gaussian  field $h(u)$
\begin{equation}\begin{array}{l}<G(\tilde{{\cal A}},u,u^{\prime};{\bf p})>=
i\int_{0}^{\infty}d\tau \exp(i\tau\frac{1}{2}({\bf
p}^{2}+m^{2}))\int_{q(0)=u} {\cal
D}q(.)\exp\Big(\frac{i}{2}\int_{0}^{\tau}(\frac{dq}{ds})^{2}\Big)\cr\times\exp(-\frac{1}{2}
\int_{0}^{\tau}ds\int_{0}^{\tau}ds^{\prime}<V(q(s))V(q(s^{\prime}))>)\delta(
u^{\prime}-q(\tau)),
\end{array}\end{equation}so that the path $q(s)$ starts at $u$ and
ends in $u^{\prime}$.

 Let us note the identity for $q(s)$ (in the sense of
the equality of expectation values; a consequence of the
transition function $p(s;u)$ in eq.(36))
\begin{equation}
q(s)=\sqrt{\tau}q(\frac{s}{\tau}).
\end{equation}
Then, in eq.(45) with the correlations (44) we have
\begin{equation}\begin{array}{l}<G(\tilde{{\cal A}},u,u^{\prime};{\bf p})>=
i\int_{0}^{\infty}d\tau \exp(i\tau\frac{1}{2}({\bf
p}^{2}+m^{2}))\int_{q(0)=\sqrt{\tau}u} {\cal
D}q(.)\exp\Big(\frac{i}{2}\int_{0}^{1}(\frac{dq}{ds})^{2}\Big)\cr\exp(-\frac{1}{2}
\tau^{2(1-\gamma)}(p_{+}p_{-})^{2}\kappa^{2}
\int_{0}^{1}ds\int_{0}^{1}ds^{\prime}<g({\bf q}(s))-{\bf
q}(s^{\prime}))> \delta(u^{\prime}-\sqrt{\tau}{\bf q}(1)).
\end{array}\end{equation}
The kernel $K_{\tau}(u,u^{\prime};{\bf p})$ (41) can be expressed
in terms of the Brownian bridge $Q$\cite{simon}( its Feynman
version is derived in \cite{hababook}) defined as the Gaussian
process on the interval $[0,1]$ ($Q(0)=0$ and $Q(1)=0$) with the
covariance ($s,s^{\prime}\geq 0$)
\begin{equation}
<Q(s)Q(s^{\prime})>=is(1-s^{\prime})\theta(s^{\prime}-s)+is^{\prime}(1-s)\theta(s-s^{\prime}),
\end{equation}
 where $\theta$ is the Heaviside step function. Let us denote by $d\nu(Q)$ the Gaussian measure with the
 covariance (48). Then,
\begin{equation}\begin{array}{l}
K_{\tau}(u,u^{\prime};{\bf p})=(2i\pi \tau)^{-\frac{1}{2}}
\exp\Big( \frac{i}{2\tau}\vert
u-u^{\prime}\vert^{2}\Big)\exp(\frac{i\tau}{2}({\bf
p}^{2}+m^{2}))\cr \int d\nu(Q)\exp\Big(i\tau
p_{+}p_{-}\int_{0}^{1}dsh(u^{\prime}s+(1-s)u+\sqrt{\tau}Q(s))\Big)
\end{array}\end{equation}
The expectation value of the propagator (47) is
\begin{equation}\begin{array}{l}<G(\tilde{{\cal A}};u,u^{\prime};{\bf p})>=
i\int_{0}^{\infty}d\tau (2i\pi \tau)^{-\frac{1}{2}} \exp\Big(
\frac{i}{2\tau}\vert
u-u^{\prime}\vert^{2}\Big)\cr\exp(i\tau\frac{1}{2}({\bf
p}^{2}+m^{2}))\int d\nu(Q)\exp\Big(-\frac{1}{2}\tau^{2(1-\gamma)}
(p_{-}p_{+})^{2}\kappa^{2}
\int_{0}^{1}ds\int_{0}^{1}ds^{\prime}\cr \times
g(\frac{1}{\sqrt{\tau}}(u^{\prime}-u)(s-s^{\prime})+Q(s)-Q(s^{\prime}))\Big).
\end{array}\end{equation}In eqs.(47)and (50) we applied the scaling
property of $g(u)$ (44).

At the end of this section let us  return to the expansion (35)
(or (39)) and calculate it with the correlation function (44)
\begin{equation}\begin{array}{l}
\int_{q(0)=u} {\cal
D}q(.)\exp\Big(\frac{i}{2}\int_{0}^{\tau}(\frac{dq}{ds})^{2}\Big)
\int_{0}^{\tau}ds_{2}\int_{0}^{s_{2}}ds_{1}<V_{s_{2}}V_{s_{1}}>+...\Big)\psi(q(\tau))
\cr
=\kappa^{2}(p_{+}p_{-})^{2}\int_{0}^{\tau}ds_{2}\int_{0}^{s_{2}}ds_{1}
p(s_{1};u-u_{1})p(s_{2}-s_{1};u_{2}-u_{1})\cr\times
p(\tau-s_{2};u_{3}-u_{2})\vert
u_{2}-u_{1}\vert^{-4\gamma}\psi(u_{3})du_{3}+....\end{array}\end{equation}
The integral over $u_{2}-u_{1}$ exists as the Lebesgue integral
only if $\gamma<\frac{1}{4}$ (for $\gamma=\frac{1}{4}$ we have a
logarithmic divergence). We encounter the same problem when
calculating (perturbatively) the expectation value of the
exponential in eq.(47). Such integrals could possibly be
interpreted in the sense of generalized functions \cite{gelfand}
allowing an extension to $\gamma> \frac{1}{4}$ (the ground state
correlation function (17) corresponds to $\gamma=\frac{1}{4}$).
Such an extension could allow to define the s-integrals in
eqs.(50)-(51). However, some positivity properties of the
integrals may be lost what can lead to difficulties in an
interpretation of the result in the framework of quantum field
theory.

An explicit calculation of the propagator (50) can be done in a
perturbation expansion in $\kappa$. The result is equivalent to
the calculation by means of the Dyson series (35). When we set
$u=u^{\prime}$ then we can obtain from eq.(50) some
non-perturbative estimates. Let us change the proper time variable
as $\tau=\tau^{\prime}(\kappa
p_{+}p_{-})^{-\frac{1}{(1-\gamma)}}$. Then, at $u=u^{\prime}$ in
the exponential in eq.(50) if $0<\gamma<1$ the term ${\bf
p}^{2}\tau= {\bf p}^{2}\tau^{\prime}(\kappa
p_{+}p_{-})^{-\frac{1}{(1-\gamma)}}$ becomes small in comparison
to the term $\tau^{\prime 2(1-\gamma)} $  in the limit of  large
$p_{z}$ (or $p_{y}$). Performing the $\tau^{\prime} $ integral
(with the negligence of the ${\bf p}^{2}\tau$ term) we obtain
\begin{equation}\begin{array}{l}<G(\tilde{{\cal A}};u,u;{\bf p})>=
C (p_{+}p_{-})^{-\frac{1}{2(1-\gamma)}}
\end{array}\end{equation}
with a certain constant $C$ (this constant can be an infinite
renormalization constant if $ \gamma\geq \frac{1}{4}$ because of
the divergence of the integral (51)). If there is no stochastic
metric ($\kappa=\gamma=0$ in eq.(50) ) then by means of the
integration over $\tau$ we obtain the propagator of the scalar
field at equal time $u=u^{\prime}$ as
\begin{equation}
({\bf p}^{2})^{-\frac{1}{2}}.
\end{equation}
We can see that if $1>\gamma>0$ then the propagator (52) is
decaying faster for large $p_{z}$ than the free propagator (53).

 Although
the propagator in a stochastic metric (44) is not Lorentz
invariant it is instructive  to calculate the Lorentz invariant
propagators with a non-canonical scaling as discussed in
\cite{graviton}\cite{becker}
\begin{equation}\begin{array}{l}
G(\xi)=i\int dp_{0}d{\bf
p}\exp(i\xi^{\mu}p_{\mu})\int_{0}^{\infty}d\tau (i\tau)^{\alpha}
\exp(i\frac{\tau}{2}p^{2}) \cr=(2\pi)^{-2} \int_{0}^{\infty}d\tau
(i\tau)^{-2+\alpha} \exp(\frac{i}{2\tau}\xi^{2})
\end{array}\end{equation}
where $p^{2}=p_{0}^{2}-{\bf p}^{2}$ and $\xi^{2}=t^{2}-{\bf
x}^{2}$. From eq.(54) the propagator in the momentum space is
$(p^{2})^{-1-\alpha}$ and the propagator at $t=0$ is $({\bf
p}^{2})^{-\frac{1}{2} -\alpha}$. Hence, in order to obtain the
behavior (52) of the equal time propagator (50) for a large
$p_{z}$ we need $(1-\gamma)^{-1}=1+2\alpha$.

\section{Summary}
We have derived an asymptotic behavior in the momentum space of
the scalar field two-point correlation function (the propagator)
in particular  Gaussian Lorentz non-invariant states. The states
describe a plane wave moving in the $x$ direction whose metric has
singular short distance correlation functions. As a result of an
interaction of the scalar field with this quantum  fluctuation of
the metric the scalar field propagator decays faster in the
momentum space (in direction orthogonal to the wave motion) than
this is possible in a canonical field theory. The canonical field
theory allows a faster decay of correlations at large distances
(small momenta) but not at small distances ( large momenta). From
the method of derivation of the result, it can be seen that such
an anomalous behavior cannot arise from a coupling of the scalar
field to a quantum gauge field or to another  scalar field, but is
characteristic to the coupling of the metric to the kinetic part
of the scalar field Lagrangian. Another special feature of the
model is that we did not calculate the scalar field correlations
in the ground state of the metric field but in a particular
time-dependent Lorentz non-invariant solution of the Schr\"odinger
equation. It is possible that  in a Lorentz invariant quantum
theory of matter fields interacting with quantum gravity the
scalar field expectation values in a time-dependent states can
decay faster in the momentum space than this is allowed in vacuum
states (if such states exist at all in theories with  quantum
gravity). In this way there remains the prospect of a construction
of quantum field theory including gravity where matter field
correlation functions in time dependent states are more regular at
short distances than this is allowed in the vacuum states of
Lorentz invariant
 theory (according to the K\"allen-Lehmann representation). Such a scheme
would be a realization of the regularizing role of Wheeler's
"quantum foam" as reviewed recently in \cite{carlip2}.


\begin{thebibliography}{99}
\bibitem{deser}S. Deser, Rev. Mod.Phys.{\bf 29},417(1957)
\bibitem{habaplb}Z.Haba, Phys.Lett.{\bf B528},129(2002)


\bibitem{jurk}J. Ambjorn, J. Jurkiewicz and R.Loll,

Phys.Rev.Lett.{\bf 95},171301(2005)
\bibitem{horava} P. Horava, Phys.Rev.Lett.{\bf 102},161301(2009)


\bibitem{reuter}M.Reuter and F. Saueressig, JHEP 1112,012(2011)
\bibitem{carlip}S. Carlip, Class.Quant.Grav.{\bf 34},193001(2017)
\bibitem{carlip2}S. Carlip, arXiv:2209.14282

\bibitem{horava2}P. Horava, Phys.Rev.{\bf D79},084008(2008)


\bibitem{verlinde}H.L. Verlinde and E.P. Verlinde,

 Nucl.Phys.{\bf
B371},246(1992)

\bibitem{kabat}D. Kabat and M. Ortiz,
Nucl.Phys.{\bf B388},570(1992)
\bibitem{detection} B.P. Abbott et al, Phys.Rev.Lett.{\bf
116},061102(2016)
\bibitem{wilczek}M. Parikh, F.Wilczek and G. Zahariade,


Phys.Rev.{\bf D104},046021(2021)

\bibitem{kuchar} K. Kuchar, J.Math.Phys.{\bf 11},3322(1970)
\bibitem{hartle} J.B. Hartle, Phys.Rev.{\bf D29},2730(1984)
\bibitem{japan}Y.Ema, R.Janno and K. Nakayama, JCAP 09(2020)015

\bibitem{inv}R. d'Inverno, Introducing Einstein's
Relativity,

Clarendon Press,Oxford,1996
\bibitem{gw}N. Christensen, Rep.Progr.Phys.{\bf
82},016903(2019)
\bibitem{ford1} H.L. Ford, Phys.Rev.{\bf D51},1692(1995)
\bibitem{ford2} H.L. Ford and N.F. Svaiter, Phys.Rev.{\bf
D54},2640(1996)
\bibitem{ford3} H. Yu, and H.L. Ford, Phys.Rev.{\bf
D60},084023(1999)

\bibitem{weinberg} S. Weinberg, Phys.Rev.{\bf 138},B988(1965)


\bibitem{sidorov} L.P. Grishchuk and Y.V. Sidorov, Phys.Rev.{\bf
D42},3413(1990)
\bibitem{albrecht}A. Albrecht, P. Ferreira, M.Joyce

and T. Prokopec,Phys.Rev.{\bf D50},4807(1994)
\bibitem{ginibre}J.Ginibre, in Statistical Mechanics and Quantum
Field Theory,edited by C. de Witt and R. Stora, Gordon and Breach,
New York,1971
\bibitem{simon}B. Simon, Functional integration and quantum
physics,

Academic, New York,1979
\bibitem{gw1}E.S. Phinney, arXiv:astro-ph/0108028,
\bibitem{gw2}P.D. Lasky et al, Phys.RevX, {\bf 6},011035(2016)




\bibitem{hababook}Z. Haba, Journ.Phys. {\bf A27},6457(1994)

\bibitem{gelfand}I.M.Gelfand and G.E. Shilov, Generalized
Functions, Vol.1, AMS, New York,1964
\bibitem{graviton}A. Bonanno,
T.Denz, J.M. Pawlowski

and M.Reichert,SciPostPhys,{\bf 12},001(2022), arXiv:2102.02217



\bibitem{becker}D. Becker and M. Reuter, JHEP12(2014)025
 \end{thebibliography}
\end{document}